# Collision dominated, ballistic, and viscous regimes of terahertz plasmonic detection by graphene

Yuhui Zhang and Michael S. Shur [a)]

*Department of Electrical, Computer and Systems Engineering, Rensselaer Polytechnic Institute, Troy, NY 12180 USA*

[a)] Author to whom any correspondence should be addressed. E-mail: shurm@rpi.edu

**ABSTRACT**

The terahertz detection performance and operating regimes of graphene plasmonic field-effect transistors (FETs) were investigated by a hydrodynamic model. Continuous wave detection simulations showed that the graphene response sensitivity is similar to that of other materials including Si, InGaAs, GaN, and diamond-based FETs. However, the pulse detection results indicated a very short response time, which favors the rapid/high-sensitively detection. The analysis on the mobility dependence of the response time revealed the same detection regimes as the traditional semiconductor materials, i.e. the non-resonant (collision dominated) regime, the resonant ballistic regime, and the viscous regime. When the kinematic viscosity ($v$) is above a certain critical viscosity value, $v_{NR}$, the plasmonic FETs always operates in the viscous non-resonant regime regardless of channel length ($L$). In this regime, the response time rises monotonically with the increase of $L$. When $v < v_{NR}$, the plasmonic resonance can be reached in a certain range of $L$ (i.e. the resonant window). Within this window, the carrier transport is ballistic. For a sufficiently short channel, the graphene devices would always operate in the non-resonant regime regardless of the field-effect mobility, corresponding to another viscous regime. The above work mapped the operating regimes of graphene plasmonic FETs, and demonstrated the significance of the viscous effects for the graphene plasmonic detection. These results could be used for the extraction of the temperature dependences of viscosity in graphene.

## I. INTRODUCTION

Ever since the discovery, graphene has been the subject of great attention due to its unique electrical [1, 2], mechanical [3, 4], chemical [5], and thermal [6] properties. One peculiar feature of graphene is that the electron-phonon scattering in graphene is weak while the electron-electron collision is very frequent due to a small effective mass [7-10]. This promotes hydrodynamic behavior and enables resonant plasmonic excitations in graphene samples [11].

Graphene-based plasmonic devices are promising for THz detection and optoelectronic applications, as they offer long-lived, high-velocity plasmons, and are highly tunable by gating and doping [11-15]. Other merits of graphene device include the significant THz absorption [16], the ability of forming multi-layer structures [17], and the high sensitivity to chemicals and vapors [5, 18]. In those devices, the realization of plasmonic resonance is a key issue. Experimental study showed that both the broadband (non-resonant) [19, 20] and resonant operation [21] were achievable in graphene plasmonic field-effect transistors (FETs), and these two detection modes are interconvertible via operating parameters.

Another important issue regarding the graphene plasmonics is the viscosity effect. Due to strong electron-electron scattering, the shear viscosity in graphene is on the order of ~0.01-0.1 m$^2$/s, which is larger than the viscosity of honey [9, 22, 23]. The electron flow in graphene could therefore behave like classical viscous liquid and even lead to the formation of whirlpools [8, 9, 24]. Consequently, the detection performance of graphene plasmonic devices will be affected. Our previous theory and measurements [25] showed that a high viscosity leads to the attenuation of DC voltage response, thus limiting the sensitivity of the plasmonic THz detector. This is caused by an additional attenuation mechanism provided by the viscosity, which becomes non-negligible when the characteristic time $L^2/v$ is close to or smaller that the momentum relaxation time, where $L$ is the channel length, and $v$ is the kinematic viscosity. [25]. The viscosity effect also broadens the plasmonic resonant peaks by $\Delta\omega_n \approx vq_n^2$, where $q_n = \pi n/L$ is the $n$th order wave number [26]. Under the pulse detection mode, a high viscosity results in a rapid decay of plasma waves and causes the saturation of the response time [27-30].

The aforementioned works revealed several important features of viscous plasmonic detection and laid the foundation of viscous detection regime. However, those studies mostly focused on other material systems such as InGaAs and Si, while the viscosity effect in graphene plasmonic FETs was rarely studied or ignored [21]. Moreover, the criteria and mechanisms of operating mode transitions in graphene plasmonic devices have not been established. Therefore, in this work, we focus on the detection of THz continuous wave (CW) and ultrashort pulse by monolayer graphene (MLG) and bilayer graphene (BLG) plasmonic FETs. The detection performance in both CW and pulsed conditions are simulated and analyzed. We also evaluate the dynamic boundaries for different detection regimes (collision dominated, ballistic, and viscous), and discuss the effect of viscosity on the mode conversion and response time alteration. Those efforts help map the operating regimes of graphene plasmonic FETs and specify the criteria of the regime conversion. We also discuss the possibility of the viscosity parameter extraction using the obtained results.

## II. BASIC PARAMETERS AND EQUATIONS

### A. Basic parameters for MLG and BLG

For a sufficiently long plasma channel, MLG and BLG follow a linear dispersion law [31-33], $\omega_p = sk$, where $\omega_p$ and $s \approx (e|U_0|/m^*)^{0.5}$ are the plasma frequency and velocity, respectively, and $k$ is the wave vector, $U_0$ is the DC gate bias above threshold. The carrier confinement and density of states are different in MLG and BLG. For MLG, the "fictitious" effective mass $m^*$ is a function of 2D carrier density ($n_s$) given by $m^* = (\hbar/V_F)\cdot(\pi n_s)^{0.5}$, where $V_F$ is the Fermi velocity on the order of ~$10^6$ m/s [8, 31, 34]. As a result, $\omega_p \sim U_0^{1/4}$. For BLG, the carrier effective mass is a much weaker function of $n_s$ and one can assume $m^* \approx 0.036 m_0$, where $m_0$ is the mass of free electron, thus $\omega_p \sim U_0^{1/2}$ [21, 33]. This shows that a BLG device has a better tuneability by gate than a MLG device.

Now we set up the mobility ($\mu$) and viscosity ($\nu$) model for MLG and BLG systems. According to previous reports [35-39], the mobility in graphene varies with carrier density $n_s$, and the $n_s$ dependence law is drastically different in MLG and BLG. $\mu$ decreases quasi-exponentially with rising $n_s$ (or $U_0$) in MLG [37-39], but it rises with an increase of $n_s$ in BLG [36]. Besides, while the general variation tendencies are similar, the value of $\mu$ in different graphene samples could vary significantly, ranging from ~0.01 m$^2$/Vs to ~20 m$^2$/Vs [7, 35, 37, 39-42]. However, the ultra-high mobility level was most realized under some strict conditions, e.g. cryogenic temperature (below 4 K), almost-intrinsic sample (near charge neutrality point), ultra-clean condition, encapsulation, and complex deposition and epitaxy procedures [37, 41, 42]. In this work, we follow the general $n_s$ dependence law proposed previously, and adjust the amplitude of $\mu$ into a typical range (0.1-0.5 m$^2$/Vs) [12, 18, 33, 36, 38, 40]. Such mobility setting may be arbitrary, but it helps us get a general variation trend and the typical order of responsivity in graphene FETs. The mobility models used are given in Fig. 1(a). As seen, when $n_s < 10^{12}$ cm$^{-2}$ the MLG mobility is higher than that of BLG, and at higher $n_s$ the mobility of BLG exceeds MLG.

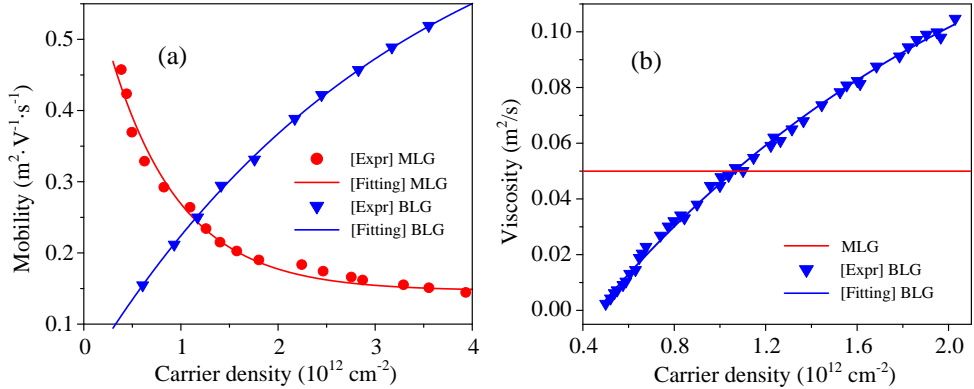

**Fig. 1.** The carrier density dependent models of (a) mobility and (b) viscosity used in our simulation at $T$ = 300 K. In (a), the experimental data (scatters) are taken from Zhu *et al.* PRB 80, 235402 (2009) [36] and multiplied with certain factors (0.6 for MLG and 5 for BLG) to fit in the common mobility range (0.1-0.5 m$^2$/Vs). The solid curves are fittings. In (b), the experimental viscosity data for BLG (blue scatters) are taken from Bandurin *et al.* Science 351, 1055-1058 (2016) [9] and converted from 240 K to 300 K assuming $1/[T^2\ln(T/T_F)]$ law [23, 43]. The solid curve is the fitting. In the simulation, we use fitting curves to calculate mobility and viscosity. The fitting equations are: $\mu_{MLG} = 0.147+0.488\exp(-n_s/0.721\times 10^{12})$ (m$^2$/Vs), $\mu_{BLG} = 0.0298+2.222\times 10^{-17}n_s-2.99\times 10^{-34}n_s^2+1.703\times 10^{-51}n_s^3$, $\nu_{BLG} = -0.0637+1.624\times 10^{-17}n_s-6.582\times 10^{-34}n_s^2+1.316\times 10^{-50}n_s^3$.

For the kinematic viscosity $\nu$, earlier works [9, 22, 23] also demonstrated different $n_s$ dependence for MLG and BLG. For MLG, the viscosity has a relatively weak $n_s$ dependence, and at $T$ = 300 K, $\nu \approx 0.05$ m$^2$/s and keeps almost unvaried as $n_s$ alters [9, 23]. For BLG, $\nu$ rises with the increase of $n_s$ [9]. In our study, we use $\nu = 0.05$ m$^2$/s at 300 K for MLG, and calculate the BLG viscosity by fitting the temperature-converted experimental data, as shown in Fig. 1(b). We can see that when $n_s <$ $10^{12}$ cm$^{-2}$, the viscosity of MLG is higher than that of BLG.

**B. Hydrodynamic regime and model**

The hydrodynamic regime in graphene has been extensively investigated in previous works [8, 10, 24, 44]. In general, to ensure the device operates at the hydrodynamic (or quasi-hydrodynamic) regime one need at least $1/\tau_{ee} > \max[1/\tau_{e\text{-ph}}, 1/\tau_{e\text{-imp}}]$ [44], where $\tau_{ee}$, $\tau_{e\text{-ph}}$ and $\tau_{e\text{-imp}}$ represent the electron-electron, electron-phonon and electron-impurity scattering times, respectively. The calculations [44] suggested that this condition is easier to be met in MLG at 300 K, while for BLG the hydrodynamic window is limited to low $n_s$ and narrows with the decrease of the operating temperature, $T$. However, the recent experiments [21] reported on the plasmonic resonance at $T = 10$ K and $T = 77$ K in a relatively large dynamic range. In our simulation, we limit the range of $n_s$ to $0.5 \times 10^{12}$ cm$^{-2}$ $< n_s < 2 \times 10^{12}$ cm$^{-2}$, and only consider the plasmonic detection under 300 K and 77 K (except for the mode validation section). The lower limit of $n_s$ is to prevent the effect of the distortion near the charge neutrality point (CNP), where the hydrodynamic theory could also fail [9, 45]. We assume that the graphene devices operating in the above parameter range fall into the hydrodynamic window.

To simulate the gated graphene plasmonic FETs, we use a one-dimensional hydrodynamic model. The governing equations consist of continuity equation, momentum relaxation equation, and energy relaxation equation, as given below [25, 46]:

$$\frac{\partial n_s}{\partial t} + \nabla \cdot (n_s \boldsymbol{u}) = 0 \tag{1}$$

$$\frac{\partial \boldsymbol{u}}{\partial t} + (\boldsymbol{u} \cdot \nabla)\boldsymbol{u} + \frac{e}{m^*}\nabla U + \frac{\boldsymbol{u}}{\tau} - \nu \nabla^2 \boldsymbol{u} = 0 \tag{2}$$

$$\frac{\partial \theta}{\partial t} + \nabla \cdot (\theta \boldsymbol{u}) - \frac{\chi}{C_v}\nabla^2 \theta - \frac{m^* \nu}{2 C_v}\left(\frac{\partial u_i}{\partial x_j} + \frac{\partial u_j}{\partial x_i} - \delta_{ij}\frac{\partial u_k}{\partial x_k}\right)^2$$
$$= \frac{1}{C_v}\left(\frac{\partial W}{\partial t}\right)_c + \frac{m^* \boldsymbol{u}^2}{C_v \tau} \tag{3}$$

The notations used in Eq. (1)-(3) are summarized and explained in Table I. We use a unified charge control model (UCCM) [25, 47] to relate $n_s$ and $U$, as presented in Eq. (4). When $eU \gg K_B T$, the UCCM reduces to $n_s = CU/e$.

$$n_s(U) = \frac{C \eta K_B T}{e^2} \ln\left(1 + \exp\left(\frac{eU}{\eta K_B T}\right)\right) \tag{4}$$

Table II. Summary of symbols used in Eq. (1)-(3)

| Symbol | Meaning | Comment |
|---|---|---|
| $\boldsymbol{u}$ | Hydrodynamic velocity (m/s) | / |
| $U$ | Gate-to-channel potential (V) | $U = U_0 - U_{ch}$<br>$U_0$: DC gate bias above threshold;<br>$U_{ch}$: channel potential |
| $\tau$ | Momentum relaxation time (s) | $\tau = \mu m^*/e$ |
| $\theta$ | Temperature (eV) | $\theta = k_B T$ |
| $\chi$ | Normalized heat conductivity (m$^2$/s) | $\chi = \kappa/n_s$<br>$\kappa$: heat conductivity |
| $C_v$ | Thermal capacitance (1) | $C_v = (\partial \Sigma/\partial \theta)_n$<br>$\Sigma$: average internal energy, $\Sigma = \theta F_1(\xi)/F_0(\xi)$<br>$\xi$: chemical potential, $\xi = \ln(\exp(E_F/k_B T)-1)$<br>$E_F$: Fermi energy<br>$F_k(\xi)$: $k$-order Fermi integral |
| $W$ | Total energy (eV) | / |
| $(\partial W/\partial t)_c$ | The collision term of $\partial W/\partial t$ | $(\partial W/\partial t)_c = (\partial \Sigma/\partial t)_c - m^* \boldsymbol{u}^2/\tau$ |

| | | (eV/s) | |

As regards the boundary condition, we use the traditional open-drain condition, i.e. $U(0, t) = U_0 + U_a(t)$ and $J(L, t) = 0$ [48, 49]. Here $U_a(t)$ is the radiation-induced small-signal voltage. For the continuous wave (CW) radiation, we assume $U_a(t) = V_{am} \cdot \cos(2\pi f t)$, where $V_{am}$ is the amplitude of incident AC voltage. For the pulse detection mode, we consider the single square pulse case, and thus the incident small-signal voltage is expressed by $U_a(t) = V_{am}(u(t)-u(t-t_{pw}))$, where $u(t)$ is the unit step function, $t_{pw}$ is the pulse width. More detailed descriptions of the hydrodynamic equations and boundary conditions can be found in [25, 46, 50].

## C. Analytic theories

Apart from the numerical modelling, we also use analytical theories to study the behaviors of graphene plasmonic FETs. Under the CW detection mode, the analytic theory predicts a constant DC voltage component between source and drain under the open drain boundary condition [48, 51]:

$$\frac{dU}{U_0} = \frac{1}{4} \frac{V_{am}^2}{U_0^2} f(\omega) \tag{5}$$

Here $dU$ is the source-to-drain voltage response, $f(\omega)$ is a frequency dependent function and is always positive. Eq. (5) shows that $dU$ is proportional to the power of incident radiation (see also Fig. 2)

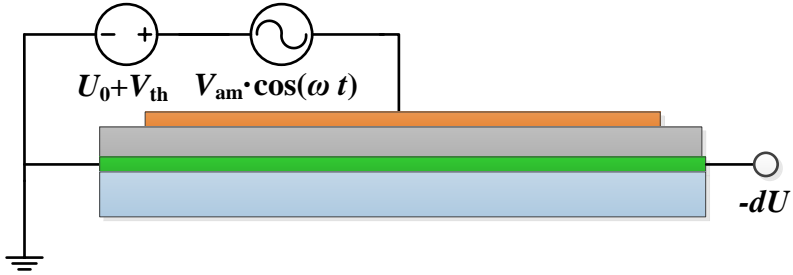

**Fig. 2**. Responsivity calculation scheme for continuous wave excitation. $V_{th}$ is the threshold voltage, $V_{am}$ is the amplitude of AC small signal voltage.

For the pulsed regime, the response performance can be evaluated by solving the linearized hydrodynamic equations in the form of $U(x, t) = \sum_{(n)} A_n \exp(\sigma_n t) f_n(x)$, where $\sigma_n$ is given by [25, 27, 28]

$$\sigma_n^{\pm} = \frac{1}{2}[-(\frac{1}{\tau} + \frac{\pi^2 v n^2}{4L^2}) \pm \sqrt{(\frac{1}{\tau} + \frac{\pi^2 v n^2}{4L^2})^2 - \frac{\pi^2 s^2 n^2}{L^2}}] \tag{6}$$

$$n = 1, 3, 5, ...$$

Here, $n$ is the odd harmonic index. If $\sigma_n$ is real, $U(x, t)$ will have a pure exponential decay waveform, corresponding to the non-resonant regime. If $\sigma_n$ has an imaginary part, $U(x, t)$ behaves an oscillatory exponential decay, and the plasmonic resonance is reached. Under the first-order approximation, the response time of the plasmonic oscillation can be defined by $\tau_r = 1/\text{Re}(|\sigma_1^+|)$.

## III. RESULTS AND DISCUSSIONS

### A. Model validation

The hydrodynamic model was used in various previous works and showed a relatively good agreement with the experimental measurements [25, 28, 51-54]. Here we compare the result of our model with the experimental data of graphene plasmonic detection, as shown in Fig. 3. Fig. 3(a1) and Fig. 3(a2) present the experimental detection responsivity using BLG FETs under two temperatures [9], and Fig. 3(b1) and Fig. 3(b2) illustrate the corresponding simulation results under similar operating conditions. As shown in Fig. 3(a1), at 10 K, $f = 0.13$ THz a relatively smooth $R_a$ curve is observed, and the absolute amplitude of $R_a$ reduces with rising $U_0$. This is a typical broadband (non-resonant) operation performance. In Fig. 3(b1) where simulation curves are given, we can see that the simulated responsivity at 0.13 THz has a similar trend as the experimental data, and the order of magnitude of $R_a$ in these two figures is the same. When the incident frequency rises to 2 THz, the experimental response curve becomes rugged, reflecting various resonant peaks. The same feature can be observed in the simulated $R_a$ curve shown in Fig. 3(b1), and the positions of resonant peaks are very close to the 2 THz experimental curve. This indicates that our hydrodynamic model can fairly well reproduce the THz detection response. A similar conformity is also obtained at 77 K, as shown in Fig. 3(a2) and Fig. 3(b2). Those results suggested the qualitative validity of the hydrodynamic model in analyzing the graphene-based plasmonic THz detection problems.

It is worth noting that the resonant peaks in Fig. 3 are relatively broad and weak. This corresponds to a broadening effect of viscosity proposed in previous papers [25, 26]. The resonant peaks might become sharper and more distinguishable with the viscosity reduction and/or an increase of the channel length.

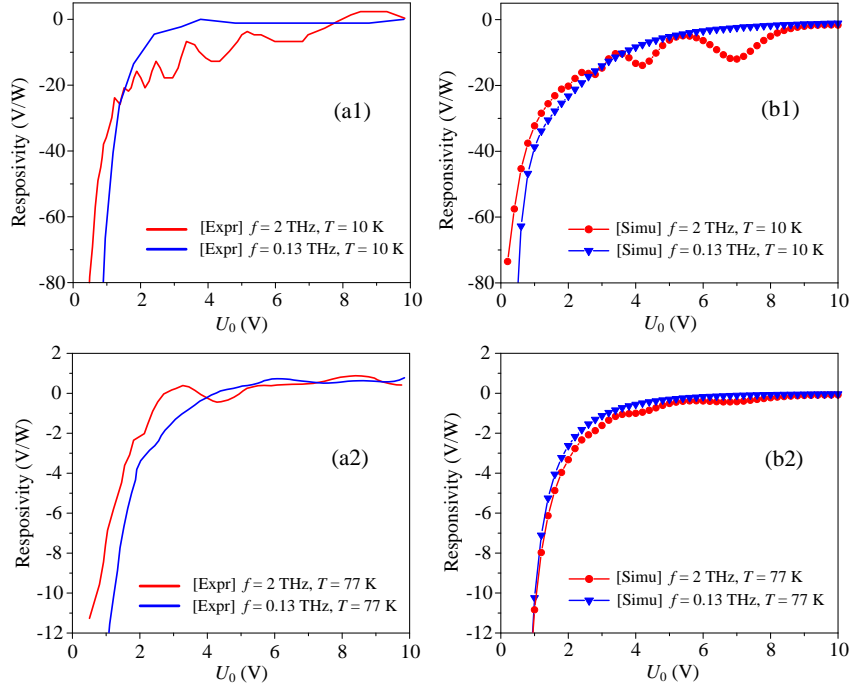

**Fig. 3**. Experimental data ((a1) and (a2)) versus the modelling results ((b1) and (b2)). The experimental data are extracted from Bandurin *et al*. Nat. Commun. 9, 5392 (2018) [21]. The operating conditions are: $L$ = 3-6 μm, $\mu \approx$ 10 m$^2$/Vs for $T$ = 10 K and $\mu \approx$ 8 m$^2$/Vs for $T$ = 77 K; the effective plasmon lifetime [21] $\tau_p \approx$ 0.3-0.9 ps; barrier layer thickness $d_b$ = 80 nm, made with hBN. The parameters used in simulation are: $L$ = 5.1 μm, $\tau_p$ = 0.8 ps for $T$ = 10 K and $\tau_p$ = 0.4 ps at $T$ = 77 K; barrier layer: $d_b$ = 80 nm, $\varepsilon_r$ = 3.8; viscosity $v$ = 0.1 m$^2$/s. The responsivity $R_a$ is calculated through $R_a = dU/P$, where $dU$ is the source-to-drain DC voltage, $P$ is the incident radiation power.

## B. Continuous wave and pulse detection characteristics

Fig. 4 presents the normalized voltage response ($R$) of graphene plasmonic FETs as functions of radiation frequency $f$ and gate bias $U_0$ under continuous wave detection mode. The normalization is done by $R = dU \cdot U_0/V_{am}^2$, where $dU$ is the absolute source-to-drain DC voltage component, $V_{am}$ is the amplitude of incident AC voltage. As seen, the amplitude of $R$ is in the order of $10^{-3} \sim 1$. Compared to our previous results [29], the voltage response level of graphene plasmonic detector is lower than that of Si, III-V and diamond devices. This is because graphene always has a small effective mass, and the mobility values we used are not extraordinary. In high mobility samples, the detection sensitivity could be enhanced. For example, if the mobility is elevated 10 times, the peak $R$ would reach ~$10^1$, which is higher than the peak $R$ of Si and III-V plasmonic FETs, but still lower than diamond devices. Besides, the simulated response data slightly deviate from the analytical curve, an effect of high viscosity [25, 29]. Due to a larger viscosity, the deviation in MLG is larger than that of BLG. Those results indicate that graphene plasmonic FET has no obvious advantage as a CW THz detector compared to FETs fabricated with other materials.

Now we evaluate the detection performance of graphene in the pulsed regime. Here we consider the single square pulse condition. Fig. 5 illustrates the response time ($\tau_r$) of MLG and BLG FETs versus the incident pulses width ($t_{pw}$) at $T = 300$ K and $T = 77$ K. As seen, the variation tendency of $\tau_r$ in graphene is identical to that in Si, III-V and diamond materials in our previous paper [30]. That is, under the short pulse condition ($t_{pw} \ll L/s$), the response time is relatively short and almost independent of $t_{pw}$. As $t_{pw}$ increases and approaches $L/s$, $\tau_r$ starts to increase, and then stabilizes when reaching the long pulse mode ($t_{pw} \gg L/s$). It is worth noting that under the operating condition given in Fig. 5, the system always operates at plasmonic resonant mode (the ballistic mode). The plasma wave travels through the channel, gets reflected, and decays with time due to friction and viscosity, thus generating an oscillatory decay voltage at the drain [27, 28, 30]. The response time in this mode can be expressed by $\tau_r = 2\tau/(1+\pi^2 v\tau/4L^2)$ [27], and $\tau_r \approx 2\tau$ under low viscosity condition. The analytical response expression is valid at the long pulse mode, while in the short pulse case the response time is reduced to approximately $\tau_r(L/2s\tau)^{0.5}$ [30]. From Fig. 5, we can see that the response time at $t_{pw} \gg L/s$ is significantly lower than $2\tau$, which indicates a strong viscosity effect. Besides, the BLG response time is longer than that of MLG due to a larger effective mass. More importantly, due to the small effective mass and high viscosity, the pulse response time in graphene is much smaller than the response time of Si, III-V and diamond materials (~0.1-1 ps) [30]. For higher mobility samples, the response time of graphene plasmonic FETs will be elevated, but we can still maintain a relatively short response time in a large mobility range (e.g. $\mu < \sim 2$ m$^2$/Vs). This suggests that the graphene plasmonic FET might be more suitable as a rapid femtosecond pulse detector than plasmonic FETs fabricated using other materials. When $T$ drops to 77 K, the graphene response time reduces, as shown in Fig. 5(b). This is mainly due to a significant increase in viscosity. Besides, the long pulse response time deviates further from the $2\tau$ limit at 77 K.

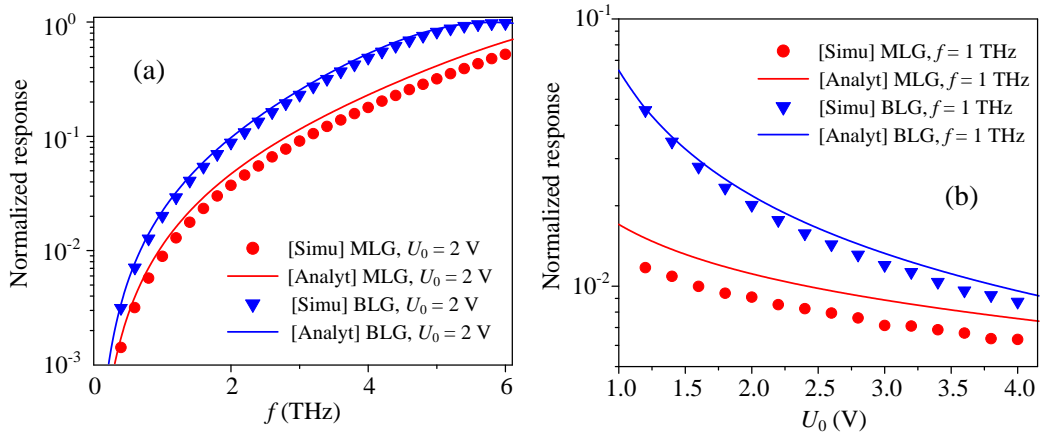

**Fig. 4.** Normalized voltage response ($R$) of MLG and BLG plasmonic FETs as functions of $f$ and $U_0$. The operating conditions are: $L = 130$ nm, $T = 300$ K, $V_{am} = 5$ mV, $d_g = 50$ nm, $\varepsilon_r = 3.8$. In (b), $U_0$ begins with 1 V to avoid being too close to the CNP.

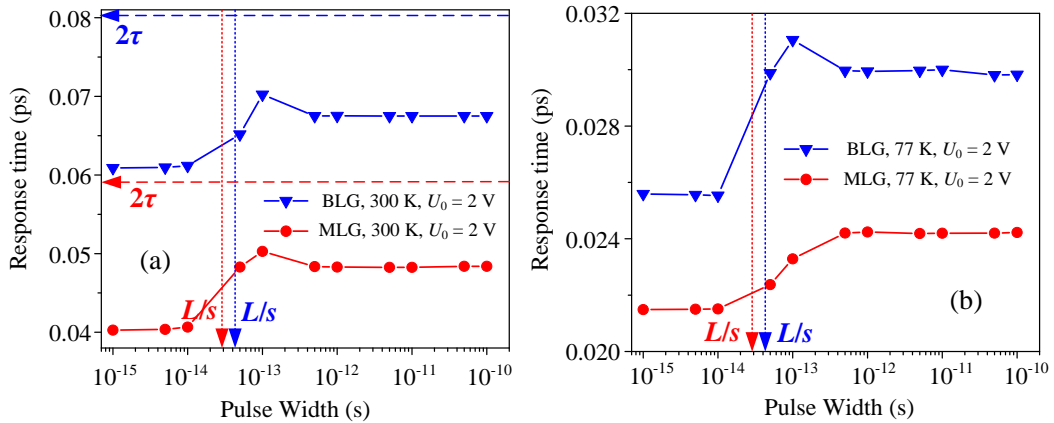

**Fig. 5.** Simulation results of response time versus the pulse width for MLG and BLG plasmonic FETs under (a) 300 K and (b) 77 K. The horizonal dashed arrow lines illustrate the $2\tau$ limit, and the dotted vertical arrow lines highlight the position of characteristic time $L/s$. The mobility: (a) 0.30 m$^2$/Vs for MLG and 0.20 m$^2$/Vs for BLG at 300 K. (b) 0.39 m$^2$/Vs for MLG and 0.17 m$^2$/Vs for BLG at 77 K. The viscosity: (a) 0.05 m$^2$/s for MLG and 0.034 m$^2$/s for BLG at 300 K. (b) 0.384 m$^2$/s for MLG and 0.263 m$^2$/s for BLG at 77 K.

## C. 3 Detection regimes and resonant boundaries

As seen from Fig. 5 the response performance of graphene FETs is highly affected by the viscosity. Since the above cases correspond to the resonant detection mode, we further study the detection properties under other regimes and further explore the viscosity effects. Fig. 6 presents the response time as a function of carrier mobility with different feature sizes, taking BLG as the example. By varying the mobility, the plasmonic FETs can be driven into different operating regimes.

When $L > 7$ nm, the $\tau_r$ curves demonstrate three distinct regions: the non-resonant (collision dominated) region, the ballistic region, and the viscous region [27, 28]. At a low mobility (i.e. $\mu < \mu_{TR}$, where $\mu_{TR}$ is a transition mobility), the plasma wave is non-exist and the device operates in the non-resonant mode. In this mode, the carrier transport is dominated by collisions with phonons and/or impurities, and the temporal voltage response is a pure exponential decay [27]. Once the mobility reaches the critical value, $\mu_{TR}$, the plasma waves are excited, and the device goes into the resonant mode. The transition mobility $\mu_{TR}$ is given by $\mu_{TR} = (Le/\pi ms)/(1-\pi v/4Ls)$, which is obtained by letting the square root term of $\sigma_1^+$ to be zero (see Eq. (6)). Under the resonant condition, if the mobility is not very high, or $\pi^2 v\tau/4L^2 \ll 1$, we have $\tau_r \approx 2\tau$ [27]. We call this region the ballistic region since the electron-electron scattering takes over but is still not strong enough to induce the viscous behavior. If the mobility is very large so that $\pi^2 v\tau/4L^2 \gg 1$, $\tau_r \approx 8L^2/\pi^2 v$, which is associated with $v$ but independent of $\mu$. Then the plasmonic FET enters the viscous regime.

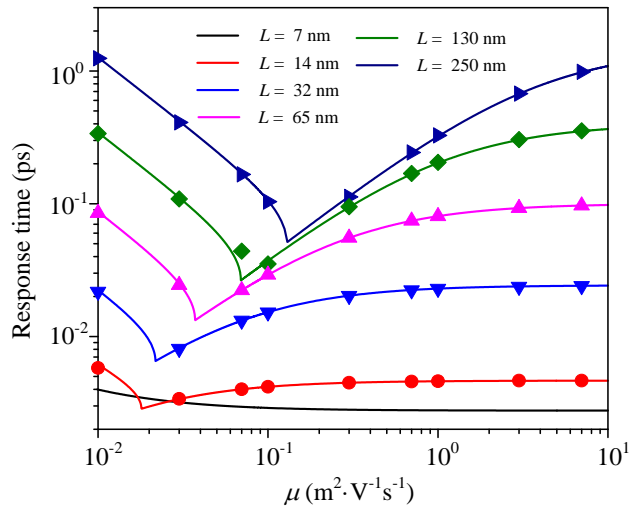

**Fig. 6.** Response time versus mobility $\mu$ for BLG plasmonic FETs with different channel lengths. The symbols represent the simulation values, and the solid lines are the analytic curves. The operating parameters are: $U_0 = 2$ V, $v = 0.034$ m$^2$/s, $T = 300$ K.

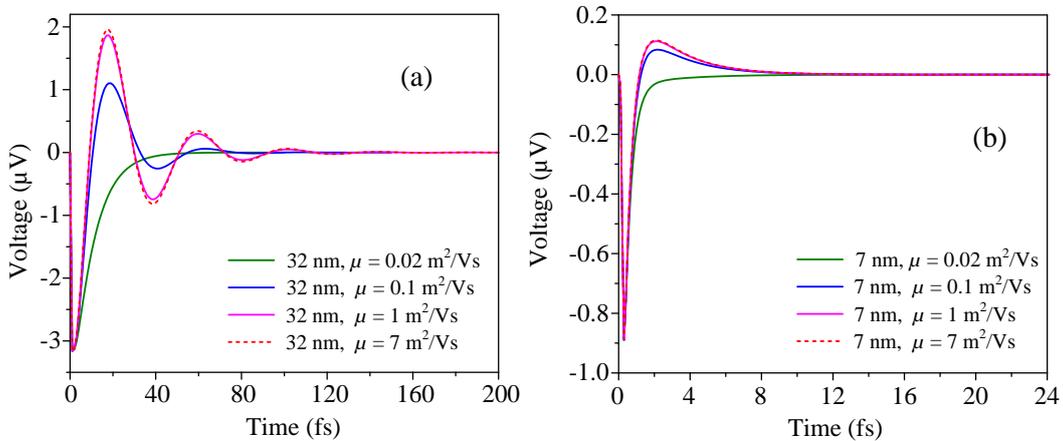

**Fig. 7.** Temporal source-to-drain voltage response of BLG plasmonic FETs at (a) $L = 32$ nm and (b) $L = 7$ nm under 4 different mobility values.

From Fig. 6, we find that the ranges of those 3 regions vary with $L$. As $L$ decreases from 250 nm to 14 nm, the critical mobility $\mu_{TR}$ decreases, and the boundary of viscous region moves leftwards. Thus, the range of viscous region expands while the band of ballistic region shrinks. As expected, the viscosity effect becomes more significant as $L$ scales down. When $L$ decreases to 7 nm, the response performance of graphene FETs changes dramatically. As seen in Fig. 6, the response time at $L$=7 nm decreases monotonically with the increase of $\mu$ and gradually saturates at large $\mu$. The simulation result also reveals a very different response profile. Fig. 7. compares the temporal voltage response waveform at $L$ = 32 nm and $L$ = 7 nm. As seen in Fig. 7(a), when $L$ = 32 nm the response waveform turns from an exponential decay (olive curve) to a quasi-sinusoidal oscillatory decay (blue, magenta and dashed red curves) as $\mu$ increases, suggesting the transition from non-resonant mode to resonant mode. In the case of $L$ = 7 nm, however, the oscillatory decay does not show up even at very high mobilities. Instead, the voltage response turns from negative to positive in few femtoseconds, and decays quasi-exponentially, exhibiting non-resonant features. The distorted temporal waveform makes us hard to calculate the response time, but we do observe the convergence of response waveform at high mobilities, corresponding to the saturation of $\tau_r$ in the analytical curve shown in Fig. 6.

The results in Fig. 6 and Fig. 7 revealed another non-resonant window at a very low channel length. To evaluate this issue in a more systematic manner, we focus on the linearized analytic theory. Note that the existence of the resonance is determined by the square root term $f(L, \tau, s, n)$ in Eq. (6):

$$f(L,\tau,s,n) = (\frac{1}{\tau} + \frac{\pi^2 v n^2}{4L^2} - \frac{\pi s n}{L})(\frac{1}{\tau} + \frac{\pi^2 v n^2}{4L^2} + \frac{\pi s n}{L}) \tag{7}$$

Using the first-order approximation and defining:

$$f_1(L,\tau,s,1) = (\frac{1}{\tau} + \frac{\pi^2 v}{4L^2} - \frac{\pi s}{L}) \tag{8}$$

we can see that $f_1$ determines the polarity of the square root term in $\sigma_1^+$. Solving $f_1 < 0$ (to ensure that $\sigma_1^+$ has an imaginary part) in terms of $\mu$ yields $\mu = (Le/\pi m s)/(1-\pi v/4Ls)$, which is the definition of $\mu_{TR}$.

We now evaluate $f_1 < 0$ in terms of $L$. Taking $x = 1/L$ as the unknown, the roots of equation $f_1 = 0$ are:

$$x_1 = \frac{2}{\pi v}(s - \sqrt{s^2 - \frac{v}{\tau}}), \quad x_2 = \frac{2}{\pi v}(s + \sqrt{s^2 - \frac{v}{\tau}}) \quad (\text{for } s^2 > \frac{v}{\tau}) \tag{9}$$

Or, equivalently:

$$L_1 = \frac{\pi \tau}{2}(s + \sqrt{s^2 - \frac{v}{\tau}}), \quad L_2 = \frac{\pi \tau}{2}(s - \sqrt{s^2 - \frac{v}{\tau}}), \tag{10}$$

The solution of $f_1 < 0$ will be discussed under the following conditions:

1) when $s^2 > v/\tau$, $L_1$ and $L_2$ are real. To ensure $f_1 < 0$ we need **$L_2 < L < L_1$**. Therefore, the range of $L$ for resonant operation is bounded at a given $\mu$. The $\Delta L$ range is given by

$$\Delta L = L_1 - L_2 = \pi \tau \sqrt{s^2 - \frac{v}{\tau}}) \tag{11}$$

We can see that $\Delta L$ increases with rising $\tau$, $s$, and decreases with the increase of $v$.

When $\tau$ (or $\mu$) is very large, $L_1$, $L_2$ and $\Delta L$ converge to:

$$\lim_{\tau \to \infty} L_1 = \pi s \tau = \lim_{\tau \to \infty} \Delta L,$$

$$\lim_{\tau \to \infty} L_2 = \lim_{\tau \to \infty} \frac{\pi v}{2(s + \sqrt{s^2 - \frac{v}{\tau}})} = \frac{\pi v}{4s} \tag{12}$$

That is, the upper limit $\lim_{\tau \to \infty} L_1 = \pi s \tau$ rises with rising $\mu$, while the lower limit $\lim_{\tau \to \infty} L_2$ is **a fixed value independent of $\mu$**.

This indicates that **when $L$ is very low, the plasmonic FETs would always operate in non-resonant mode regardless of the mobility.** Also, the upper limit $\pi s \tau$ is in fact equivalent to the boundary for broadband detection ($s\tau/L \ll 1$) proposed in previous papers [48]. This means that the plasma oscillation generated near the source side cannot reach the drain side due to a long channel. It is also noted that the lower limit $L = \pi v/4s$ is a pole in the expression of $\mu_{TR}$. With a relatively large viscosity, this lower limit of $L$ in graphene is larger than that in other traditional materials. Hence, it is easier to achieve the mode conversion by scaling in graphene.

2) when $s^2 < \nu/\tau$, or $\nu > s^2\tau \approx \mu U_0$, we have $f_1 > 0$. This means at a given mobility, there exists a critical viscosity $\nu_{NR} \approx \mu U_0$ so that **when $\nu > \nu_{NR}$, the plasmonic FET cannot reach the resonance mode regardless of the channel length**. Under this circumstance, $\text{Re}[\sigma_1^+] = \sigma_1^+$, so the response time can be expressed by

$$\tau_r = \frac{1}{(|\sigma_1^+|)} = \frac{2}{|-(\frac{1}{\tau}+\frac{\pi^2\nu}{4L^2})+\sqrt{(\frac{1}{\tau}+\frac{\pi^2\nu}{4L^2})^2-\frac{\pi^2 s^2}{L^2}}|} \qquad (13)$$

$$= 2[(\frac{L^2}{\pi^2 s^2\tau}+\frac{\nu}{4s^2})+\sqrt{(\frac{L^2}{\pi^2 s^2\tau}+\frac{\nu}{4s^2})^2-\frac{L^2}{\pi^2 s^2}}]$$

This expression is valid for all non-resonant cases. When $L \Rightarrow 0$, $\tau_r \Rightarrow \nu/s^2$. At large $L$, $\tau_r \sim \frac{4L^2}{\pi^2 s^2\tau}+\frac{\nu}{s^2}$, i.e. the response time rises quadratically with the increase of $L$.

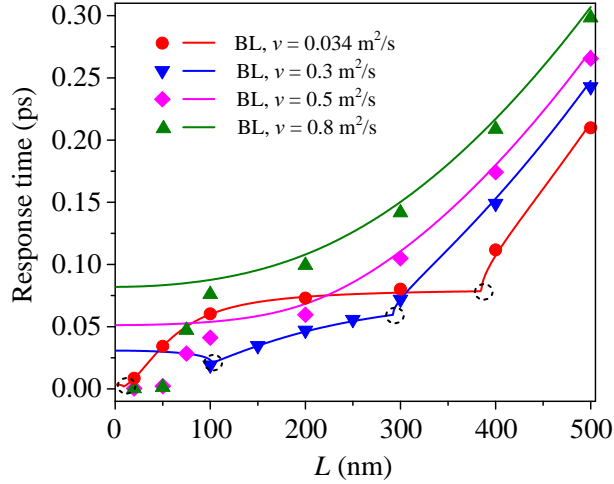

**Fig. 8.** Response time versus $L$ for different viscosity values for BLG FETs at $U_0 = 2$ V, $t_{pw} = 10^{-11}$ s. Simulation values are presented by symbols, and the analytic results are given by solid curves. The critical viscosity $\nu_{NR} = 0.392$ m²/s. The dashed circles highlight the boundaries of non-resonant/resonant regions.

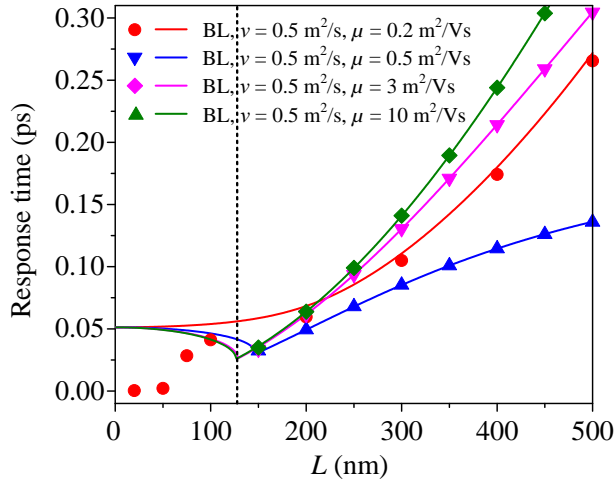

**Fig. 9.** Response time versus $L$ at different mobilities for BLG FETs under $U_0 = 2$ V, $t_{pw} = 10^{-11}$ s, $\nu = 0.5$ m²/s > $\nu_{NR}$. The dashed line highlights the position of $L = \pi\nu/4s$.

To evaluate the validity of above theory, Fig. 8 compares the simulation and analytic results of $\tau_r$ for BLG plasmonic FETs as a function of $L$. As seen, at low viscosity ($\nu < \nu_{NR}$), the resonant band can be observed in the analytic curve, and the simulation values agree well with the analytic results. With the increase of viscosity, the resonant band narrows, as predicted by Eq. (12). When $\nu$ exceeds $\nu_{NR}$, the resonant band disappears, and the system operates at a fully non-resonant regime. The simulation data exhibit a good agreement with the analytic curves for relatively large $L$ cases. When $L$ < 100 nm, the modeled $\tau_r$ values drop below the analytic curve, and approach 0 when $L$ < 20 nm, indicating the failure of analytic theory. From the simulation, we noticed that the plasma wave velocity $s$ under $L$ < 100 nm is higher than the theoretical value. Therefore, the decrease of response time at low $L$ might be caused by the increased $s$ and thus reduced $v/s^2$. The speed-up of plasma wave here could be related to the viscosity effect. A more in-depth mechanism of this phenomenon is still missing and requires further investigation. In the design of plasmonic FETs, we may use this phenomenon to further shorten the response time and accelerate the device response.

Fig. 9 shows a high viscosity BLG FET ($v$ = 0.5 m²/s) at different mobilities. As before, the simulated response time data agree with the analytic curves except those at short channel region. When the mobility rises, the system can be driven out of fully non-resonant mode (since $v_{NR} = \mu U_0$ rises). However, the lower resonant boundary cannot cross the limit $L = \pi v/4s$, since it's independent of $\mu$.

Now we evaluate the boundary of viscous region. As mentioned earlier, to ensure the domination of viscosity effect, we need, from Eq. (6):

$$\frac{1}{\tau} \leq \frac{\pi^2 v}{4L^2} \Rightarrow v \gg v_{cr} = \frac{4L^2}{\pi^2 \tau} \tag{14}$$

where $v_{cr}$ is a characteristic viscosity. For $v \ll v_{cr}$, the loss is dictated by scattering; for $v \gg v_{cr}$, the loss is determined by viscosity (viscous regime). Rewriting this inequality into the following form:

$$v \geq \alpha v_{cr} = \frac{4\alpha L^2}{\pi^2 \tau_{Tv}} \Rightarrow \tau_{Tv} \geq \frac{4\alpha L^2}{\pi^2 v} \Rightarrow \mu_{Tv} \geq \frac{4\alpha e L^2}{\pi^2 v m} \quad (\alpha \gg 1) \tag{15}$$

We obtain $\mu_{Tv}$, the approximate mobility boundary of viscous regime. Apparently $\mu_{Tv}$ drops with the decrease of $L$, suggesting the leftward expansion of viscous region. This confirms to our observation in Fig. 6. In general, $\mu_{Tv} \gg \mu_{TR}$, but when $L$ is very low, $\mu_{Tv} < \mu_{TR}$, and from this inequality we obtain:

$$\mu_{Tv} < \mu_{TR} \Rightarrow L < L_{cr} = \frac{\alpha + 1}{4\alpha} \frac{\pi v}{s} \tag{16}$$

It can be seen that the characteristic length $L_{cr} \approx \pi v/4s$, which is exactly the lower boundary of resonant region. This result indicates that in a short channel device, the carrier transport is dominated by the viscosity. One possible reason is that the mean free path of electron-phonon scattering or electron-impurity scattering becomes comparable or even larger than the channel length under this condition, and thus these scattering rates are restricted and viscosity takes over. With a strong viscosity and a short channel, the carriers cannot reach the resonant oscillation.

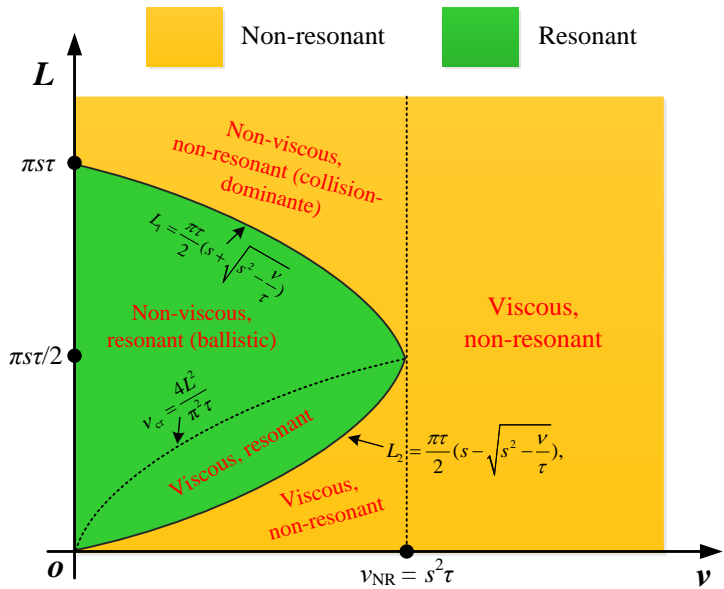

**Fig. 10.** Schematic of operating regimes in a graphene plasmonic FET.

Fig. 10 summarizes the variation of operating regimes in a $L$-$v$ scale. As seen, the viscous regime penetrates into both resonant and non-resonant modes, suggesting the significant role of viscosity in graphene plasmonic detection.

Table II. Material parameters and $L_1$, $L_2$, $\Delta L$, $v_{NR}$ in for different materials

| Materials | $U_0$ (V) | $\mu$(m$^2$/Vs) | $v$ (m$^2$/s) | $m^*/m_0$ | $s$ (m/s) | $\tau$ (fs) | $L_2$ (nm) | $L_1$ (nm) | $\Delta L$ (nm) | $v_{NR}$ (m$^2$/s) |
|---|---|---|---|---|---|---|---|---|---|---|
| MLG | 2 | 0.3 | 0.05 | 0.017 | 4.54×10$^6$ | 29.17 | 8.85 | 406.8 | 397.9 | 0.6 |
| BLG | 2 | 0.2 | 0.034 | 0.036 | 3.13×10$^6$ | 39.92 | 8.79 | 383.2 | 374.4 | 0.4 |
| Si | 2 | 0.1 | 2.35×10$^{-4}$ | 0.19 | 1.36×10$^6$ | 10.80 | 0.14 | 461.7 | 461.5 | 0.2 |
| GaN | 2 | 0.2 | 1.47×10$^{-3}$ | 0.23 | 1.24×10$^6$ | 26.16 | 0.93 | 1015.3 | 1014.3 | 0.4 |
| InGaAs | 2 | 1 | 0.04 | 0.041 | 2.93×10$^6$ | 23.32 | 10.78 | 2134.5 | 2123.7 | 2.0 |
| p-diamond | 2 | 0.3 | 1.6×10$^{-4}$ | 0.663 | 7.28×10$^5$ | 11.31 | 0.17 | 2587.9 | 2587.7 | 0.6 |
| n-diamond | 2 | 0.5 | 5.17×10$^{-4}$ | 0.36 | 9.88×10$^5$ | 10.24 | 0.41 | 3178.0 | 3177.6 | 1.0 |

Note: the material parameters are taken from: [31-33, 36-39] (MLG and BLG), [55-58] (Si), [58-60] (GaN), [27, 61, 62] (InGaAs), [63-66] (p- and n-diamond).

Table II shows the typical material parameters and $L_1$, $L_2$, $\Delta L$, and $v_{NR}$ for graphene, Si, GaN, InGaAs and diamond materials. As seen, the MLG and BLG have relatively large $L_2$ and relatively small $L_1$ values compared to those of other materials. Note that the mobility and viscosity of graphene is adjustable in a large dynamic range, thus $L_1$, $L_2$, $\Delta L$, and $v_{NR}$ can also be adjusted by changing $\mu$ and $v$. For example, in a high viscosity sample, the lower limit $L_1$ can increase to ~100 nm, as shown in Fig. 9; given a large mobility sample, the upper limit $L$ could rise to several μm. This shows that graphene has a much better mode tunability compared to other materials.

**D. Viscosity extraction**

As reported in our previous paper [30], the viscosity of the plasmonic FET material can be extracted using $\tau_r = 2\tau/(1+\pi^2 v\tau/4L^2)$ in the resonant mode. That is, by plotting $\tau_r^{-1}$ vs $L^{-2}$ and get $v$ from the slope of the curve (since $1/\tau_r = 1/2\tau + \pi^2 v/8L^2$). Now we discuss the viscosity extraction at the non-resonant mode. For $s^2 > v/\tau$, the analytic expression of $\tau_r$ is given by Eq. (13). This equation predicts that when $L \Rightarrow 0$, $\tau_r \Rightarrow v/s^2$. However, simulation results in Fig. 8 and Fig. 9 show that the analytic expression is invalid at low $L$. Since the region where the analytic theory fails begins at around $L = s\tau$ (~125 nm in our BLG FET case), and the analytic curve of $\tau_r$ becomes very flat in this region, we could assume $\tau_r(L=s\tau) \approx v/s^2$, and therefore $v \approx s^2\tau_r(L=s\tau)$. A more rigorous method is to fit the experimental data under relatively large $L$. The fitting curve follows equation (13) given by

$$\tau_r(L,v) = 2[(C_1 L^2 + C_2 v) + \sqrt{(C_1 L^2 + C_2 v)^2 - C_3 L^2}] \tag{17}$$

Here, $C_1 = 1/\pi^2 s^2\tau$, $C_2 = 1/4s^2$, $C_3 = 1/\pi^2 s^2$. To obtain $C_1$, $C_2$ and $C_3$, we need to know $\mu$ ($\tau = \mu m/e$) and $s$. The plasma velocity $s$ can be obtained from the delay time data.

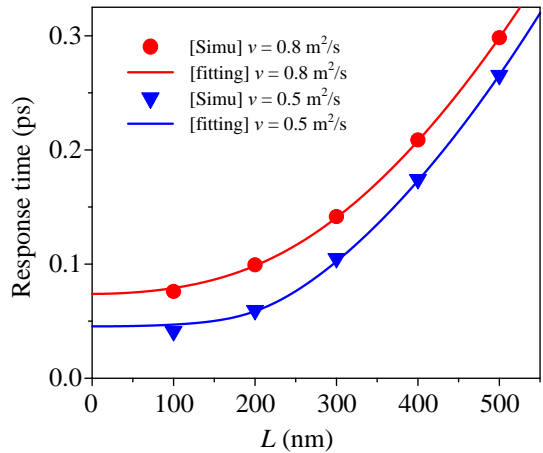

Fig. 11 Simulation points and fitting curves (following equation (17)) of response time for BLG FETs under $v = 0.5$ m$^2$/s and $v = 0.8$ m$^2$/s.

Fig. 11 shows the fitting results of simulation data for BLG FETs under $v = 0.5$ m$^2$/s and $v = 0.8$ m$^2$/s using equation (17).

As seen, the fitted curves exhibit a good agreement with the analytical curves given in Fig. 8. The fitting yields $v = 0.724$ m$^2$/s for $v = 0.8$ m$^2$/s case and $v = 0.445$ m$^2$/s for $v = 0.5$ m$^2$/s case, suggesting that this method tends to underestimate the viscosity. Nevertheless, it still provides a relatively high accuracy for viscosity extraction. Compared to the resonant viscosity extraction method proposed earlier [30], this non-resonant method demonstrates a higher accuracy, possibly because of a higher polynomial order of the fitting function. We could use this method to evaluate the temperature dependence of viscosity in graphene.

## CONCLUSION

The THz continuous wave (CW) detection and ultrashort pulse detection by monolayer graphene (MLG) and bilayer graphene (BLG) FETs were studied by a validated hydrodynamic model and compared with the analytical theory and experimental data. The simulation revealed a much more important role of viscosity in graphene compared to more traditional semiconductor materials. Graphene response time to short excitation pulses could be much smaller than that of Si, GaN, InGaAs, or diamond FETs. Therefore, graphene and bilayer graphene FETs could be used as femtosecond pulse detectors. At relatively low viscosities, the resonant window in graphene plasmonic FETs were found to be limited in a certain channel length range, and this range shrunk with the increase of viscosity. At a sufficiently large viscosity, the resonant window disappeared, and the device would always operate on the non-resonant, viscous regime. We developed the analytical expression of response time in this regime, and it demonstrated a good agreement with the simulation data. This expression could be also used for the graphene parameter extraction including the electron and hole viscosity measurements.

## ACKNOWLEDGEMENTS

This work was supported by xxx.

## DATA AVALIBILITY

The data that support the findings of this study are available from the authors upon reasonable request.